# Stimulus competition by inhibitory interference

Paul H.E. Tiesinga
Department of Physics & Astronomy
University of North Carolina at Chapel Hill
Chapel Hill, North Carolina 27599

Phone: 919 962 7199
Fax: 919 962 0480
Email: tiesinga@physics.unc.edu

# Stimulus competition by inhibitory interference

**Abstract**

When two stimuli are present in the receptive field of a V4 neuron, the firing rate response is between the weakest and strongest response elicited by each of the stimuli alone (Reynolds et al, 1999, Journal of Neuroscience 19:1736-1753). When attention is directed towards the stimulus eliciting the strongest response (the preferred stimulus), the response to the pair is increased, whereas the response decreases when attention is directed to the other stimulus (the poor stimulus). When attention is directed to either of the two stimuli presented alone, the firing rate remains the same or increases slightly.

These experimental results were reproduced in a model of a V4 neuron under the assumption that attention modulates the activity of local interneuron networks. The V4 model neuron received stimulus-specific asynchronous excitation from V2 and synchronous inhibitory inputs from two local interneuron networks in V4. Each interneuron network was driven by stimulus-specific excitatory inputs from V2 and was modulated by a projection from the frontal eye fields. Stimulus competition was present because of a delay in arrival time of synchronous volleys from each interneuron network. For small delays, the firing rate was close to the rate elicited by the preferred stimulus alone, whereas for larger delays it approached the firing rate of the poor stimulus. When either stimulus was presented alone the neuron's response was not altered by the change in delay.

The model suggests that top-down attention biases the competition between V2 columns for control of V4 neurons by changing the relative timing of inhibition rather than by changes in the degree of synchrony of interneuron networks. The mechanism proposed here for attentional modulation of firing rate – gain modulation by inhibitory interference – is likely to have more general applicability to cortical information processing.

# Introduction

The neural correlates of selective attention have been studied in monkeys using recordings from single neurons in cortical area V4 (Connor et al., 1996; Fries et al., 2001; McAdams and Maunsell, 1999a; McAdams and Maunsell, 1999b; McAdams and Maunsell, 2000; Moore and Armstrong, 2003; Moore and Fallah, 2004; Reynolds and Chelazzi, 2004; Reynolds et al., 1999; Reynolds and Desimone, 1999; Reynolds and Desimone, 2003; Reynolds et al., 2000). A key finding is that attention modulates both the mean firing rate of a neuron in response to a stimulus and the coherence of spikes with other neurons responsive to the same stimulus (Fries et al., 2001). The increase of coherence with attention is strongest in the gamma frequency range (30-80Hz).
Networks of inhibitory interneurons (Beierlein et al., 2000; Beierlein et al., 2003; Galarreta and Hestrin, 1999; Galarreta and Hestrin, 2001; Gibson et al., 1999) have been implicated in the generation of synchronous gamma-frequency-range oscillations in the hippocampus (Fisahn et al., 1998; Whittington et al., 1995) and the cortex (Deans et al., 2001) and could entrain a large number of principal cells (Koos and Tepper, 1999; Tamas et al., 2000) as well as modulate their firing rates (Aradi et al., 2002). Hence, synchrony modulation of interneuron networks could mediate the effects of attention observed in cortical neurons (Tiesinga and Sejnowski, 2004). The degree of synchrony of the inhibitory inputs to V4 neurons can be characterized by their temporal dispersion, referred to here as precision. We proposed that attention could act by increasing the precision of inhibitory inputs to V4 neurons (Tiesinga et al., 2004). We found that the modulation of the model neuron's firing rate and its coherence with the inhibitory oscillation was consistent with the observed effects of attention. These models only considered the case of one stimulus in the neuron's receptive field.

The response of a V4 neuron to two stimuli in its receptive field has also been studied (Chelazzi et al., 1993; Gawne and Martin, 2002; Luck et al., 1997; Moran and Desimone, 1985; Reynolds et al., 1999). One of the stimuli yielded a weak response when presented alone, whereas a more vigorous response was elicited when the other stimulus was presented alone. The latter is referred to as the preferred stimulus and the former is referred to as the poor stimulus. When both stimuli were presented at the same time, the neuron's firing rate was less than the response to the preferred stimulus but larger than the response to the poor stimulus. This result is consistent with the framework of stimulus competition (Desimone, 1998) where the pair response was a weighted sum of the responses to the stimuli when presented alone. When attention was directed to the preferred stimulus the response increased, whereas attending to the poor stimulus decreased the response. Hence, attention biased the outcome of stimulus competition towards the stimulus that was attended. The neural circuit that underlies stimulus competition is not yet fully characterized. Our goal is to determine whether and how modulation of the activity of local interneuron networks can account for attentional modulation of stimulus competition. We find that modulating the relative phase of synchronized interneuron networks rather than the degree of synchrony can account for the competition between V2 columns for the control of V4 neurons. We propose that the projection from the frontal eye fields (FEF) to area V4 (Moore and Armstrong, 2003) can modulate the relative phase between interneuron networks providing inputs to the V4 cells.

# Methods

In a previously studied model of inhibitory interneurons connected by chemical synapses, the network produced oscillatory activity that consisted of a sequence of synchronized spike volleys (Diesmann et al., 1999; Tiesinga and Jose, 2000). First, we describe the statistics of the output of a local interneuron network. Each spike volley was characterized by the number of spikes in the volley $a_{IV}^i$ (with $i$ the volley index, $a$ the activity and IV indicating inhibitory volley), their mean spike time $t_{IV}^i$ and their spike-time dispersion $\sigma_{IV}^i$. The spike-time dispersion, $\sigma_{IV}$, is inversely related to the precision P. The mean number of spikes per volley, $a_{IV}$, is determined by the fraction of network neurons that is active on a given cycle, the size of the network, and the presynaptic release probability (Koch, 1999). The two interneuron networks were not explicitly simulated; rather the above statistics were used to model input spike trains representing the synchronous inhibitory input, as described below. These input spike trains will be referred to as 'network activity' throughout.

The method used to obtain synchronous volleys is as described in (Tiesinga et al., 2002). A set of volley times $t_{IV}^i$ (with a fixed intervolley interval equal to 25ms) was generated for the first network. The volley times for the second network were obtained by adding a fixed delay D to each of the first network's volley times. Next, a binned spike-time probability (STP) was obtained by convolving all the volley times with a Gaussian filter with standard deviation $\sigma_{IV}$ and area $a_{IV} \Delta t$ (the bin width $\Delta t = 0.01 ms$ was equal to the integration time-step used in the simulations). Input spike times were generated as a Poisson process from the STP, as in (Tiesinga et al., 2002). Each input spike produced an exponentially decaying conductance pulse in the postsynaptic cell yielding a current $I_{syn} = \Delta g_{inh} \exp(-t/\tau_{inh})(V - E_{GABA})$. In this expression $t$ is the time since the last presynaptic spike, $\tau_{inh} = 5ms$ is a decay time constant (Bartos et al., 2002), $\Delta g_{inh} = 0.05 mS/cm^2$ is the unitary synaptic conductance, $V$ is the postsynaptic membrane potential, and $E_{GABA} = -75mV$, is the reversal potential for GABAergic inhibitory synapses. The neuron was also driven by asynchronous excitatory synaptic inputs. The parameters were, with a notation analogous to that for inhibitory inputs, $\tau_{exc} = 2ms$, and $\Delta g_{exc} = 0.02 mS/cm^2$. The reversal potential for fast AMPA excitatory synapses was $E_{AMPA} = 0mV$ (Shepherd, 1998).

The resulting train of conductance pulses drove a single compartment neuron with Hodgkin-Huxley voltage-gated sodium and potassium channels, a passive leak current, and excitatory and inhibitory synaptic currents as described above (Wang and Buzsaki, 1996). Full model equations are given in (Tiesinga and Jose, 2000). They were integrated using a noise-adapted 2nd-order Runge-Kutta method (Greenside and Helfand, 1981) implemented in the Fortran programming language, with time step $dt = 0.01 ms$.

Simulations were run multiple times with different seeds for the random number generator, yielding different trials. Spike times $t_i^j$ ($i$th spike time during the $j$th trial) of the target neuron were calculated as the time that the membrane potential crossed 0 mV from below. The spike phase was expressed in ms and calculated as $\phi_i^j = \mod(t_i^j + 12.5, 25)$. The mean interspike interval, $\tau$, was calculated as the mean of all intervals during a given trial and then averaged across all trials. In most cases, we took one 200s long trial. The mean firing rate $f$ was $1000/\tau$ ($\tau$ is in ms, $f$ in Hz). Histograms were calculated using Matlab's (The Mathworks) *hist* function. The binwidth was 1ms for phase histograms, 2 Hz for firing rate histograms and 0.1 for the histograms of firing rate ratios.

## Results
### A model for stimulus competition

A number of different models have been proposed to account for stimulus competition and its modulation with attention. Reynolds and coworkers proposed a simple phenomenological model (Figure 1A). A pool of excitatory and inhibitory neurons is associated with the poor stimulus (pool 1) and another pool is associated with the preferred stimulus (pool 2). When a particular stimulus is presented, the corresponding pool is activated. For a preferred stimulus, the pool generates relatively more excitation than inhibition, whereas for a poor stimulus inhibition is stronger and excitation is weaker. In their model, the firing rate of the V4 neuron is proportional to the total amount of excitation it receives, divided by the sum of excitation and inhibition. Stimulus competition emerges automatically because the summed inhibition decreases the pair response more than the summed excitation increases it. Attention is assumed to increase the strength of the inputs from the pool associated with the attended stimulus. For attending the preferred stimulus that means relatively more excitation in response to the pair, hence an increased firing rate, whereas for attending the poor stimulus there would be relatively more inhibition, hence a decreased firing rate. The model accounts succinctly for the experimental observations, but its biophysical underpinnings are unclear. The firing rate cannot be written in the form that is assumed for the Reynolds model unless normalizing effects due to recurrent network activity are included (Carandini et al., 1997). The pools providing inputs to the model neuron are likely to be located in area V2. However, the projections from one cortical area to the other are thought to be primarily excitatory (Shepherd, 1998). The question then is, where does the inhibition come from? And, how are the excitatory and inhibitory inputs correlated?

<Figure 1 about here>

A basic assumption of the Reynolds model is that the amount of excitation and inhibition provided by a pool only depends on whether the associated stimulus is present. Specifically, it is not influenced by the presence of other stimuli not associated with the given pool. This is different from previous models for stimulus competition (see (Usher and Niebur, 1996) and the Discussion). Here we determine how stimulus competition in area V4 emerges from independently activated pools in V2. The following framework is used to study the behavior of the V4 model neuron (Figure 1B). There are two excitatory pools. The first pool is exclusively activated by the poor stimulus (stimulus 1) and provides weak excitation to the neuron. The second pool is exclusively activated by the preferred stimulus (stimulus 2) and provides strong excitation to the neuron. For most cases, we took a 3:1 ratio for the rates of preferred and poor excitatory inputs, respectively. This means that when both stimuli are present pool 1 provides 25% of the excitatory inputs and pool 2 provides 75%. 100% excitation corresponded to 2000 excitatory inputs per second. The input rate is proportional to the number of neurons in the pool, their mean firing rate and the probability of a successful spike transmission across the synapse. We assume that each stimulus activates the neurons in the associated pool in the same way. The neurons in the pool will therefore have comparable mean rates. The difference between the poor and preferred stimuli has to do with the number of synapses, transmission probability and/or synaptic strength. In the model, the synapses all have the same strength. Short-term plasticity and transmission failure are also not taken into account. Hence, within the framework of the model, poor and preferred stimuli are only different in the rate of excitatory inputs that they provide to the neuron.

Each excitatory pool also activates a corresponding inhibitory pool. When the inhibitory pools are activated they produce synchronized volleys with a precision P at a rate of 40Hz. The precision is the inverse of the temporal dispersion of the spikes in a volley and it is expressed in 1/ms. A more synchronous inhibitory network has a higher value for the precision. We used P values between 0 and 1/ms. The level of activation of the interneuron networks is the same for poor and preferred stimuli. Each volley had on average $a_{IV} = 18.75$ synaptic inputs, yielding 750 inputs per second for each inhibitory pool. The volleys produced by each network arrive at the neuron at different times. The delay in volley arrival times is expressed in ms and denoted by D. We hypothesize that a top-down projection from the FEF modulates the precision and phase of the inhibitory volleys. The delay is the phase difference between the two networks and it is also referred to as the relative phase. Phases are expressed as a time between 0 and 25ms rather than the more conventional choice of values between 0 and 1. Using this model system we determine how modulation of D and P can account for the observed effects of attention on stimulus competition.

<Figure 2 about here>
<Table 1 about here>

**Stimulus competition by inhibitory interference**

The firing rate response to the poor stimulus alone and to the preferred stimulus alone is denoted by $f_1$ and $f_2$, respectively. When both stimuli are presented simultaneously the response is $f_3$. Stimulus competition thus implies the following inequality: $f_1 < f_3 < f_2$. For the reader's convenience, the reported effects of attention on the different stimulus configurations are summarized in Table 1. The volleys from the second inhibitory network (activated by stimulus 2) were delayed by D=7.5ms compared with volleys from the first inhibitory network (activated by stimulus 1). The firing rate in response to stimulus 2 alone was higher than in response to stimulus 1 alone because the second excitatory pool provided a higher excitatory input rate (Figure 2A top and bottom). When both stimuli were presented simultaneously the response was less than the response to stimulus 2 alone, because the two inhibitory networks were out of phase. We hypothesized that the decrease in firing rate was due to the shortening of the time interval on each cycle during which the inhibitory conductance was small enough to allow the neuron to spike. This time interval is referred to as the spiking interval. To investigate this further the distribution of the spike phase was calculated (Figure 3). This will tell us at what point during the oscillation cycle the neuron spiked. When only one interneuron network was active the phase histogram had one peak and the neuron was only able to spike during an interval that was about half the cycle length (Figure 3A). When the excitatory drive to the neuron was doubled the spiking interval increased to about 2/3 of a cycle length. The phase histogram had also become bimodal because on some cycles the neuron spiked twice (Figure 3B). When both interneuron networks were active with a relative phase of 7.5ms, there was one peak in the phase histogram (Figure 3C), which was sharper than before. The neuron only spiked during an interval of about 1/3 of a cycle length. For a higher level of excitation, the spiking interval had increased to half a cycle length (Figure 3D). Note that there is no bimodality in the phase histogram because it is hard to fit two spikes in a single spiking interval. When the two networks were completely out of phase (D=12.5ms, Figure 3E,F) the phase histograms were bimodal from the start and the spiking was confined to two brief periods on each cycle. The minimum interspike interval is determined by the absolute and relative refractory period. In order to produce multiple spikes on a given cycle the spiking interval needs to be long enough to fit the minimum interspike interval for that neuron. Hence, the strength of the excitatory drive and the period during which the neuron can spike determine the spike rate. Stimulus competition occurs because of modulation of the spiking interval.

<Figure 3 about here>

We investigated how the firing rate for the single stimulus and pair configuration depended on the delay and precision of the interneuron networks. In the absence of inhibition, the firing rate versus excitatory input rate curve had a steep onset (Figure 4Aa). When inhibition from a single interneuron network was added there was a regime where the firing rate did not increase as fast with the excitatory input rate. This is the fluctuation-driven regime where the spikes are caused by the voltage fluctuations rather than a mean upward drift of the membrane potential (see the Discussion). The gain of the firing rate versus input rate curve was approximately the same for both cases. The fluctuation-driven regime extended over a larger range when both interneuron networks were active and the gain of the response curve was less than for the single interneuron network. Hence, inhibition does not only shift the response curves to the right, it also induces a fluctuation-driven regime and changes the gain of the response (Chance et al., 2002; Mitchell and Silver, 2003; Prescott and De Koninck, 2003). The gain of the response curves also decreased when the delay between the networks was increased (Figure 4Ab) or the precision was decreased (Figure 4Ac).

<Figure 4 about here>

Let us denote the response of a neuron with one active interneuron network by $F_1(f_e)$ and the response with two active networks by $F_2(f_e)$ with $f_e$ being the excitatory input rate. Then $f_1 = F_1(\frac{1}{4} f_e)$, $f_2 = F_1(\frac{3}{4} f_e)$ and $f_3 = F_2(f_e)$, and stimulus competition implies that $F_1(\frac{1}{4} f_e) < F_2(f_e) < F_1(\frac{3}{4} f_e)$. These three curves are plotted together in Figure 4B. We use a qualitative scale for precision: low precision corresponds to $0 \leq P \leq 1/5$, moderate precision corresponds to $1/5 < P \leq 1/3$ and high precision corresponds to $P>1/3$. When the precision is moderate to high, stimulus competition is obtained for all input rates (Figure 4Ba,b). However, for low precision and low input rates, the pair response can dip below the response to stimulus 1 alone (Figure 4Bc). This suggests that the inhibitory networks need to be synchronized in order to obtain stimulus competition.

This precision requirement was further investigated for different combinations of poor and preferred stimuli. The excitatory input rate to the neuron in response to stimulus 1 and stimulus 2 alone is denoted by $f_{e1}$ and $f_{e2}$, respectively. We determined $f_3 = F_2(f_{e1} + f_{e2})$ and the ratios $f_3/f_1 = F_2(f_{e1} + f_{e2})/F_1(f_{e1})$ and $f_3/f_2 = F_2(f_{e1} + f_{e2})/F_1(f_{e2})$ for all distinct pairs of $(f_{e1}, f_{e2})$ values. We used sixty values for $f_{e1}$ and $f_{e2}$ that were between 0 and 1475 inputs per second in increments of 25 spikes per second. Stimulus competition is obtained when the first ratio is larger than one and the second ratio is less than one. The histograms of these ratios are shown in Figure 5 together with the histogram of the values of $f_3$ for which there was stimulus competition. For a small delay, D=2.5ms, and moderate precision, P=1/(3 ms), stimulus competition was obtained for 91% of the pairs (Figure 5A). The pair responses were broadly distributed between 10 and 60 Hz. For the majority of the pairs the response was closer to $f_2$ than to $f_1$. When the delay was increased to its maximal value, D=12.5ms, less than 1% of of the pairs yielded stimulus competition (Figure 5B). This was because the inhibition was so effective that the pair response usually was below $f_1$. For low precision, stimulus competition was obtained in about 24% of the pairs (Figure 5C). The pair response was in the majority of the cases closest to $f_1$. In summary, small delays and moderate precision are required to get robust stimulus competition.

<Figure 5 about here>

**Attentional modulation by changes in synchrony**

It was recently reported that altering the precision of the inhibitory inputs to a neuron could lead to a gain change of its firing rate response curve. Increasing input precision led to an increased firing rate, whereas decreasing precision led to a decreased firing rate (Tiesinga et al., 2004). We therefore studied the following scenario for attentional bias: When both stimuli are present and attention is directed toward the poor stimulus, the precision of the inhibitory inputs to the neuron is decreased, leading to a lower firing rate that is closer to $f_1$. When attention is directed toward the preferred stimulus, the precision is increased, yielding a response closer to $f_2$. Attention is thus hypothesized to have a different effect on the synchrony of the inhibitory networks depending on whether the focus of attention is on a poor or a preferred stimulus. This type of behavior may be hard to orchestrate in cortical networks, but the aim here is to see whether in principle such a mechanism could work.

<Figure 6 about here>

First, the effect of a global inhibitory network was considered (Figure 6Aa). There was only one pool of inhibitory neurons whose precision was modulated. The response to stimulus 1 alone, stimulus 2 alone and stimulus 1 & 2 together, always increased with the precision of the global inhibitory network. However, the pair response, $f_3$, was always larger than $f_2$ and $f_1$. Hence, no stimulus competition was obtained. Next, we returned to the original model with two separate inhibitory pools (Figure 1B), but their precisions were varied independently. The precision of the second inhibitory pool was increased to model directing attention to stimulus 2. The response to stimulus 2 alone and both stimuli at the same time increased, whereas the response to stimulus 1 remained the same (Figure 6Ab). The precision of the first inhibitory pool was decreased to model attending stimulus 1. The response to stimulus 1 alone and both stimuli at the same time decreased, whereas the response to stimulus 2 remained the same (Figure 6Ac). By contrast, the experimental result is that when a single stimulus in the receptive field is attended, the neuron's response stays the same or increases moderately (McAdams and Maunsell, 1999a). This problem could be addressed by making the direction of precision change depend on whether a stimulus is presented by itself or with another stimulus. This would imply a direct interaction between the inhibitory networks, which we excluded from the model.

<Figure 7 about here>

**Attentional modulation by inhibitory interference**

We investigated whether we could model the effects of attention by varying the delay between the interneuron networks rather than by varying the precision (Figure 7). For attention directed to stimulus 1, the delay was increased and the firing rate decreased from the baseline condition (Figure 7, top panels). In contrast, for attention directed to stimulus 2, the delay was decreased and the firing rate increased from the baseline value (Figure 7, bottom panels). The attentional bias should be able to raise the firing rate to values close to $f_2$ and decrease the firing rate to values close to $f_1$. The requirements necessary to achieve this dynamic range were determined (Figure 6B). For high precision, the firing rate was strongly modulated by the value of the delay, but the pair response did not get close to $f_1$ (Figure 6Ba). For low precision, there was only a weak modulation of the firing rate with delay and the pair response remained close to $f_1$ (Figure 6Bc). Only for moderate precision, P=(1/3ms), there was both a strong modulation with delay and a pair response that went from values close to $f_1$ to values close to $f_2$ (Figure 6Bb). Hence, there is only a limited range of precision values for which attentional modulation is possible.

**Discussion**

We have proposed a single cell mechanism for stimulus competition in V4 based on temporal interference of synchronous inhibition. The key idea is that a neuron's firing rate can be modulated by changing the length of the spiking interval without needing to change the amount of excitation. The spiking interval is the length of that part of the cycle during which a neuron is free to fire and it is determined by the phase difference between different sources of synchronous inhibition. The requirements for attentional modulation of stimulus competition were as follows. (1) The feedforward excitation from V2 needs to be asynchronous. (2) The local inhibitory networks need to be synchronized. (3) Attention-related bottom-up or top-down projections need to be able to modulate the relative phase of synchronized interneuron networks. In the following we will discuss these requirements for stimulus competition, discuss other mechanisms for stimulus competition and compare gain modulation by inhibitory interference to previously proposed mechanisms for firing rate modulation.

**Requirements for stimulus competition by inhibitory interference**

*Asynchronous excitation*. Let us assume for a moment that excitation was synchronous. In that case, a change in phase of one of the interneuron networks also changes the relative phase between the synchronous excitation and inhibition the neuron receives. The relative timing of inhibition and excitation strongly affects the neuron's firing rate. Hence, changing the delay would not only affect the pair response, as intended, but also the response to a stimulus presented alone. Thus excitation needs to be asynchronous.
Usually spike trains produced by a group of neurons receiving synchronous inhibition are synchronous. Attentional modulation is thought to operate on multiple levels in the visual pathway (Luck et al., 1997; Reynolds et al., 1999). If that is the case, how can the V2 excitatory inputs be asynchronous? The first possibility is that there is no synchrony between excitatory neurons that provide inputs to V4 because the local circuitry in layer 2/3 destroys synchrony. This is possible but would impose constraints on the layer 2/3 circuit (Brunel, 2000b; Gerstner, 2000; van Vreeswijk and Sompolinsky, 1996). Another possibility is that the axonal transmission delays from V2 to V4 are sufficiently variable to wash out any synchrony that was present in V2. This would not affect the synchrony of the interneuron networks in our model because they are local to V4 and presumably located in layer 4.

*Synchronous inhibition*. Experiments (Fisahn et al., 1998; Whittington et al., 1995) and theoretical and computational work (Aradi and Soltesz, 2002; Bartos et al., 2002; Borgers and Kopell, 2003; Brunel, 2000a; Brunel and Wang, 2003; Golomb and Hansel, 2000; Neltner et al., 2000; Tiesinga and Jose, 2000; Wang and Buzsaki, 1996; White et al., 1998) show that interneuron networks readily synchronize in the gamma frequency range. Although the synchronization dynamics of inhibitory networks has been studied extensively using model simulations the focus has almost exclusively been on the stationary state, rather than dynamic changes in synchrony. Here we considered the case where interneuron networks were not active if there was no stimulus present. However, in a more realistic setting they would be asynchronous and less active without a stimulus present and synchronized and strongly active with a stimulus present. In that case, interneuron networks would need to switch rapidly between different synchrony states. We found two types of networks whose synchrony can be changed by modulatory inputs. First, in a purely inhibitory network, synchrony can be modulated by increasing excitation to a part of the network. The activated neurons increase their firing rate, synchronize and reduce the activity of the other group of interneurons. Synchrony can be modulated using this mechanism on time scales as short as 100ms (Tiesinga and Sejnowski, 2004). Second, in a mixed excitatory and inhibitory network, synchrony can be modulated by activating the interneuron network when the inhibitory and excitatory neurons are mode-locked to each other (Buia and Tiesinga, 2004). In that case, synchronized excitatory activity recruits inhibitory activity that temporarily shuts down the excitatory activity. When the inhibition decays the excitatory neurons become active again and the cycle starts over. Activation of interneuron networks by neuromodulators may increase their synchrony, in turn increasing excitatory synchrony, but without altering the mean firing rate of individual neurons (Buia and Tiesinga, 2004).

*Modulating relative phase*. In order to modulate the relative phase between the two networks one only needs to modulate the absolute phase of one of the networks. We performed model simulations of a set of synaptically coupled inhibitory and excitatory neurons similar to those discussed in the preceding paragraph. When the excitatory neurons were mode-locked to the inhibitory neurons the phase of the interneuron network could be modulated by dynamically applying a current pulse to it (Buia and Tiesinga, 2004). There are probably other network configurations in which the same could be achieved. However, the model results show that is in principle possible to change the phase between isolated interneuron networks.

*Which interneurons are involved in inhibitory interference?* There is an enormous diversity of interneurons in the cortex. It was recently reported that there are two dynamically distinct inhibitory networks in layer 4 of the somatosensory cortex (Beierlein et al., 2003). The fast-spiking (FS) cells receive inputs from the thalamus and provide feed-forward inhibition (Galarreta and Hestrin, 1999). The FS cells are an important component of feed-forward models dealing with the emergence of orientation-selectivity in primary visual cortex (Miller, 2003). The low-threshold spiking (LTS) cells are coupled by electric gap junctions and can easily become synchronized (Gibson et al., 1999). The interneurons in our model have the same stimulus-specificity as the excitatory inputs to the model neuron, but they should also synchronize easily. The model networks thus have properties in common with both the FS and LTS networks. The short-term synaptic dynamics of the two networks are drastically different and lead to distinct temporal response properties (Beierlein et al., 2003). Experiments are needed that probe the temporal dynamics of attentional modulation of stimulus competition in order to identify which of the two networks is modulated by attention.

*Which neurons provide modulatory input to V4 interneuron networks?* Moore and coworkers recently studied the neural interactions between the FEF and V4 in awake macaque monkeys (Moore and Armstrong, 2003; Moore et al., 2003). When neurons in the FEF were electrically stimulated – microstimulation – the animal made a saccade to a particular area in visual space. They then stimulated FEF with a smaller current that did not elicit a saccade and recorded at the same time from V4 neurons with a receptive field at the location of the intended target of the saccade. Microstimulation did not affect the response of the V4 neuron when there was no stimulus in its receptive field. However, when there was a stimulus in the receptive field the firing rate increased. The increase was larger for stimuli that elicited a large response without microstimulation. This resembles attentional gain modulation of orientation tuning curves in V4 as reported in (McAdams and Maunsell, 1999a). When a distractor stimulus was placed outside the receptive field, the strength of the modulation of V4 responses by the FEF increased. This is consistent with the effects of attention on the V4 response to two simultaneously presented stimuli (Desimone and Duncan, 1995; Luck et al., 1997; Reynolds et al., 1999). These experimental results provide support for the involvement in attentional processing of the postulated modulatory projection from the FEF to local interneuron networks in V4 (Figure 1B).

**Other approaches to stimulus competition**

Stimulus competition has been achieved in network models (Usher and Niebur, 1996). The model consisted of feature-selective principal cells, such as, for instance, cells sensitive to stimulus orientation. Each principal cell provided excitatory input to a global inhibitory network and in turn received inhibitory inputs from it. When multiple stimuli were presented, the response of the principal cell to a preferred stimulus was suppressed by the inhibition recruited by other principal cells responding to poor stimuli. The competition could be biased in favor of a particular orientation by top-down excitation to the corresponding principal cell. When the stimulus corresponding to the attended orientation was presented, the sum of the feed-forward and top-down inputs allowed the responsive neuron to recruit the strongest inhibition, which suppressed the neurons responding to the non-attended orientations. An extension of this model (Deco et al., 2002) was used to account for the results in (Reynolds et al., 1999). The difference between these proposals and ours is that the role of inhibitory synchrony was not studied in their models.

There are also single cell mechanisms for stimulus competition. Archie and Mel (Archie and Mel, 2000) proposed that stimulus competition arose from the spatial segregation of afferent synapses onto different regions of the excitable dendritic tree of V4 neurons. This raises the issue of whether asynchronous excitation and inhibition would be enough to obtain stimulus competition in single compartment model neurons. Neurons have two operating regimes (Kuhn et al., 2004; Tiesinga et al., 2000). A neuron can be in the fluctuation-driven regime where inhibition dominates excitation. The spikes are then due to stochastic fluctuations in the membrane potential that cross the action potential threshold. Or a neuron can be in the current-driven regime where there is a net upward drift in the membrane potential that causes the action potentials. In that case, excitation dominates inhibition and the firing rate is proportional to the difference between excitation and inhibition. Consider a neuron that is operating in the current-driven regime. A poor stimulus as well as a preferred stimulus elicit spikes and lead to a firing rate that exceeds the spontaneous rate (this is zero in the model but typically a few spikes per second in experiment). This means that excitation is larger than inhibition in both cases. When the input due to the poor stimulus is added to that due to the preferred stimulus, the firing rate has to exceed that elicited in response to the preferred stimulus alone, because the net difference between excitation and inhibition has increased. Therefore stimulus competition does not occur in the current-driven regime. Can stimulus competition occur in the fluctuation-driven regime? In the fluctuation-driven regime, the variance of the membrane potential fluctuations can decrease when the inputs due to poor and preferred stimulus are added (Burkitt et al., 2003; Chance et al., 2002; Tiesinga et al., 2000). In that case, the firing rate would also decrease and fall below the response to the preferred stimulus. This is indeed what happened for low precision inhibition in Figure 4Ac. For the model considered here, the resulting firing rate was low – the pair response was closer to the response to the poor stimulus. However, for a different model the low firing rate problem can probably be resolved. Hence, stimulus competition is possible in neurons driven by asynchronous excitatory and inhibitory spike trains, as long as they operate in the fluctuation-driven regime. The real problem is the attentional modulation. How can attention modulate the poor-stimulus–related inputs to the V4 neuron in such a way that the response increases or remains the same when the poor stimulus is presented alone, but decreases when both stimuli are presented simultaneously. A decrease in the amount of inhibition or an increase in the amount of excitation to the V4 neuron would increase its response to stimulus 1 alone, but would also increase the pair response. The same holds for increasing the precision of the inhibitory inputs. The solution suggested here is to alter the relative phase of volleys produced by the interneuron networks. When only one interneuron network is active and the excitatory inputs are asynchronous, the absolute phase of the volleys does not affect the firing rate (it does affect the timing of the spikes). However, when two interneuron networks are active, the relative phase does matter. The pair response can then be made to decrease with attention by increasing the delay. It is possible to also increase the response to stimulus 1 alone by increasing the precision of the inputs with attention. Even for that case, the pair response will decrease with increasing delay (data not shown).

**Mechanisms for gain modulation**

Firing rate modulation by interference of inhibitory inputs may have more general applicability as a mechanism for gain modulation. Previously, three other mechanisms have been proposed for how gain changes can be achieved (Burkitt, 2001; Burkitt et al., 2003; Chance et al., 2002; Destexhe et al., 2001; Doiron et al., 2001; Fellous et al., 2003; Holt and Koch, 1997; Kuhn et al., 2004 ; Larkum et al., 2004; Mitchell and Silver, 2003; Murphy and Miller, 2003; Prescott and De Koninck, 2003; Rauch et al., 2003; Salinas and Sejnowski, 2000; Tiesinga et al., 2004; Tiesinga et al., 2000; Ulrich, 2003). We briefly summarize them and discuss how they relate to gain modulation by inhibitory interference.

*Gain modulation by balanced synaptic inputs.* Under *in vivo* conditions neurons receive a constant barrage of excitatory and inhibitory inputs (Shadlen and Newsome, 1998). The synaptic inputs are called balanced when the effective reversal potential of the sum of excitatory and inhibitory inputs is equal to the neuron's resting membrane potential (leak reversal potential). By proportionally scaling the rates of excitatory and inhibitory inputs the amplitude of the voltage fluctuations can be modulated while maintaining a constant mean membrane potential. In the balanced mode the neuron is driven by fluctuations: the larger the fluctuations, the higher the firing rate (Chance et al., 2002). Interestingly, an increase in balanced activity decreased the gain of the firing rate versus current curve (Burkitt, 2001; Burkitt et al., 2003; Kuhn et al., 2004; Tiesinga et al., 2000). The saturation of dendritic nonlinearities can further enhance the change in gain obtained with balanced inputs (Prescott and De Koninck, 2003).

*Gain modulation by tonic inhibition and excitation.* Tonic inhibition by itself did not lead to multiplicative gain modulation (Doiron et al., 2001; Holt and Koch, 1997). However, when tonic inhibition was applied in combination with either excitatory or inhibitory Poisson spike train inputs, changes in gain as well as shifts in sensitivity were observed (Mitchell and Silver, 2003; Ulrich, 2003). Murphy and Miller showed that changes in tonic excitation and inhibition can lead to approximate multiplicative gain modulation of cortical responses when the nonlinearity of the thalamic contrast response is taken into account (Murphy and Miller, 2003).

*Gain modulation by correlations.* When a neuron is in a fluctuation-driven state and receives inputs from different neurons, it is sensitive to correlations between these neurons. Stronger correlations lead to an increase in the amplitude of voltage fluctuations, hence to an increase in firing rate (Salinas and Sejnowski, 2000). Gain modulation by inhibitory synchrony is a specific example of this mechanism. Changing inhibitory input synchrony resulted in a gain change for neurons receiving on the order of ten inhibitory inputs on each oscillation cycle (Tiesinga et al., 2004).

*Gain modulation by inhibitory interference.* This mechanism assumes that there are multiple sources of synchronous inhibitory inputs. When all of the input sources are in phase – no interference – the gain is high, whereas when they are not in phase – interference – the gain is low. The network proposed here is similar to circuitry that makes neurons in the inferior colliculus sensitive to interaural timing differences (Brand et al., 2002). This circuitry was studied in vitro using dynamic clamp (Grande et al., 2004). Two synchronized and periodic excitatory input trains were injected into a neuron. The excitatory volleys in each train arrived at a rate of $f_d$. The volleys of the second train were delayed with respect to the first train. For a small delay, the neuron's firing rate was $f_d$, the neuron fired one spike for every two input volleys. However, for the maximal possible delay, when the two input trains were in antiphase, the firing rate of the neuron had doubled to $2f_d$. By contrast, we find that for small delays the firing rate is higher than for longer delays.

It seems hard to modulate the activity of cortical networks in such a way that the synaptic inputs they provide to other neurons remain balanced. To the best of our knowledge no network architecture has been proposed that achieves this. By contrast, tonic excitation and inhibition can be easily modulated using neurotransmitters or modulators. The same manipulations can also alter the correlations in networks (see, for instance, (Tiesinga and Sejnowski, 2004)). The attractive feature of gain modulation by inhibitory interference is that the gain depends on the conjunction of two inputs. This bears a resemblance to the idea of binding: spikes that occur at the same phase are part of the same percept and are transmitted and processed together (for review see (Singer and Gray, 1995)). Here the same idea is used to gate information transmission via gain changes.

**Open problems and future work**

Our model was constructed to account for the results obtained by Reynolds and coworkers (Reynolds and Chelazzi, 2004; Reynolds et al., 1999). The response of V4 cells to two simultaneously presented stimuli was also studied by (Gawne and Martin, 2002) (see also the review by (Rousselet et al., 2003)). They attempted to make the distance between the two stimuli as large as possible. Under those conditions the response of the neuron was closer to the maximum of the response to either of the stimuli alone rather than a weighted average of these responses as was the case in (Reynolds et al., 1999). The model presented here can also account for these results: When the phase delay is small, the firing rate is close to the maximal response to either of the stimuli alone. These results underscore the need for a more detailed investigation of stimulus competition as a function of the distance between the stimuli as well as the difference in feature values, such as, for instance, color and orientation. In addition, characterizing the dynamics of the FEF modulation of V4 responses remains a challenging problem.

The general picture that emerges from the proposed model is a cortex with patches of active and synchronized interneuron networks that all fire at a particular phase. Bottom-up and top-down projections dynamically modulate the phase in order to preferentially process the behaviorally relevant stimuli. From a modeling standpoint this raises a number of questions. What is the typical size of a patch with neurons that fire at approximately the same phase? Does the firing phase change continuously across the cortical surface or are there discontinuous transitions? How are the phases dynamically modulated and with what time scale? These and other questions will be addressed in future work using large-scale network models.

**Figure captions**

Figure 1. The model neuron and the synaptic inputs it receives. (A) The model proposed by Reynolds and coworkers (Reynolds et al., 1999). Stimulus 1 and 2 each activate a separate pool of inhibitory and excitatory neurons. (B) Temporal Interference Network. The model neuron receives feed-forward inputs from two excitatory pools in area V2 and it receives inhibition from two local interneuron networks in V4. Each excitatory pool is associated with a specific stimulus and projects only to the corresponding interneuron network. The interneuron network is modulated by inputs from the frontal eye fields.

Figure 2. Stimulus competition emerges from the interference between two inhibitory networks. In each panel, from top to bottom, we show the response to (or input due to) the presentation of stimulus 1 alone, stimulus 1 & 2 together, and stimulus 2 alone. (A) The inhibitory (solid lines) and excitatory input rates (dashed lines) and (B) the neuron's membrane potential are plotted versus time. The input rates were 500, 1500, 750 and 750 inputs per second for the first and second excitatory pool and the first and second inhibitory pool, respectively. The other model parameters are given in the Methods section.

Figure 3. The delay between inhibitory networks determines the firing rate via modulation of the spiking interval. Phase histograms for the spike times of a V4 neuron driven by (A,B) one active interneuron network and by two active interneuron networks with a delay of (C,D) 7.5ms and (E,F) D=12.5ms. The phase histograms were calculated over 200s of data. The responses for (A) to (F) were, expressed as (excitatory input rate in Hz, output rate in Hz), (780, 32.2), (1580, 89.7), (1170, 3.3), (2370, 41.8), (1170, 0.2), (2370, 20.9), respectively.

Figure 4. The relative phase and precision of interneuron networks determine the gain of the firing rate response curves. (A) Firing rate of the V4 neuron as a function of the excitatory input rate for purely excitatory inputs (dot-dashed line), for excitatory inputs together with one active interneuron network (dashed) and together with two active interneuron networks (solid line). The parameters for (a) were the same as in Figure 2. In (b) the delays in ms were, from left to right, D=0, 7.5 and 12.5 (note that the response with one active interneuron network does not depend on the delay, therefore only one curve is shown). In (c) the precisions in 1/ms were, for the solid and the dashed lines, from left to right, P=1/3, 1/5 and 1/10. (B) Analysis of stimulus competition. Solid lines represent the neuron's response for one active interneuron network together with 25% excitation (lower curves) and 75% excitation (upper curves). The dashed lines are the responses to 100% excitation with two active interneuron networks. The rate on the x-ordinate corresponds to 100% excitation. Stimulus competition is present when the dashed curve is between the two solid curves. Parameters (D in ms, P in 1/ms) were (a) (7.5, 1/3), (b) (12.5, 1/3) and (c) (7.5, 1/10).

Figure 5. Small delays and moderate precision are necessary for robust stimulus competition. In each panel, we show the histogram of (a) $f_3/f_1$ and (b) $f_3/f_2$. In (c) we plot the distribution of $f_3$ for the pairs which satisfied the requirement for stimulus competition. Parameter values were (D in ms, P in 1/ms)= (a) (2.5, 1/3), (b) (12.5, 1/3) and (c) (7.5, 1/10).

Figure 6. Modulation of relative phase could account for attentional bias, whereas modulation of precision did not. In each panel, solid lines are the neuron's responses with a single active interneuron network together with 25% excitation (lower curves) and 75% excitation (upper curves). The dashed lines are the responses to 100% excitation with two active interneuron networks. (A) Firing rate as a function of precision. We varied the precision of (a) the global

interneuron network, the interneuron network associated with (b) stimulus 2 and (c) stimulus 1. (B) Firing rate is plotted as a function of delay with the precision (in 1/ms) equal to (a) 1, (b) 1/3 and (c) 1/8.

Figure 7. Stimulus competition is biased by modulating the relative phase of the interneuron networks. We plot (A) the inhibitory input rates and (B) the neuron's voltage in response to stimulus 1 and 2 presented simultaneously. In each panel, from top to bottom, stimulus 1 is attended, attention is directed away from the neuron's receptive field and stimulus 2 is attended. The precision was P=1/(3ms) and the delay in ms was, from top to bottom, D=12.5, 7.5, 2.5.

Table 1

| Stimulus configuration | Attended object | Response | Response change from baseline |
| --- | --- | --- | --- |
| Stimulus 1 | Stimulus 1 | $f_1$ | Same or slight increase |
| Stimulus 2 | Stimulus 2 | $f_2$ | Same or slight increase |
| Stimulus 1 & 2 | Stimulus 1 | $f_3$ | Decrease |
| Stimulus 1 & 2 | Stimulus 2 | $f_3$ | Increase |

Summary of the experimentally observed responses of V4 neurons to different stimulus configurations and under different attentional conditions. See (Reynolds and Chelazzi, 2004; Reynolds et al., 1999).

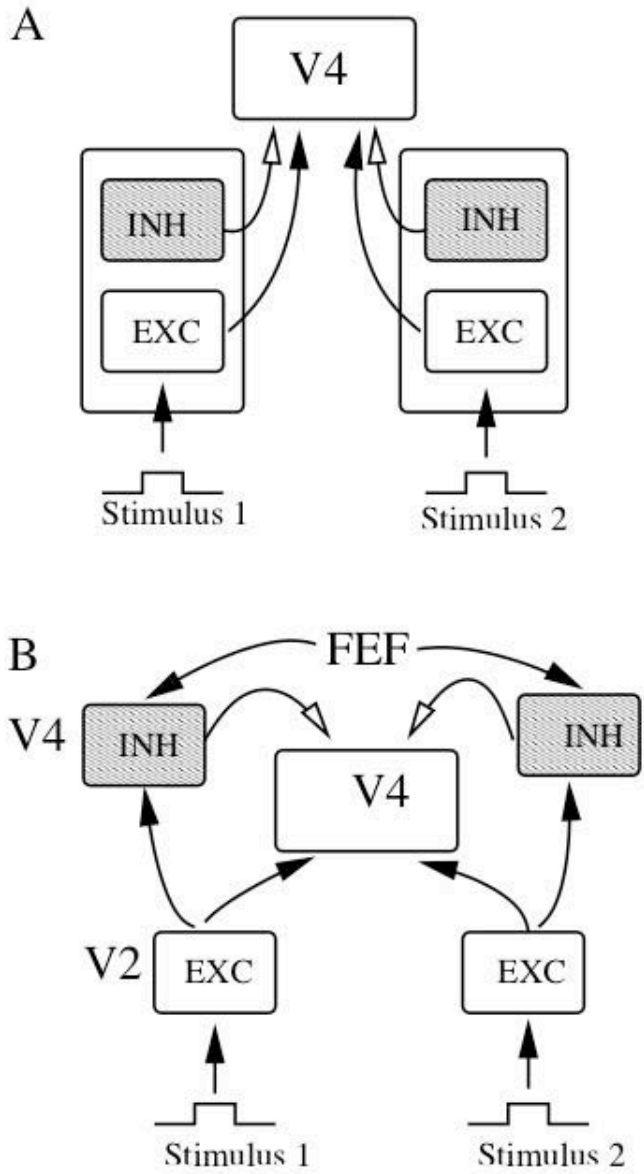

Figure 1

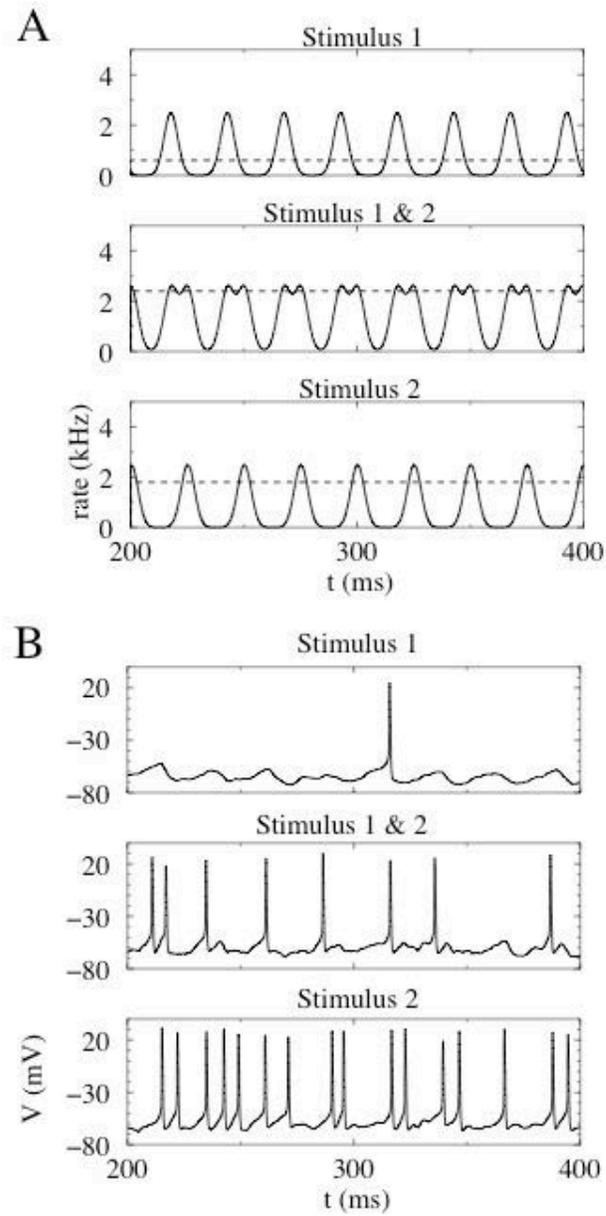

Figure 2

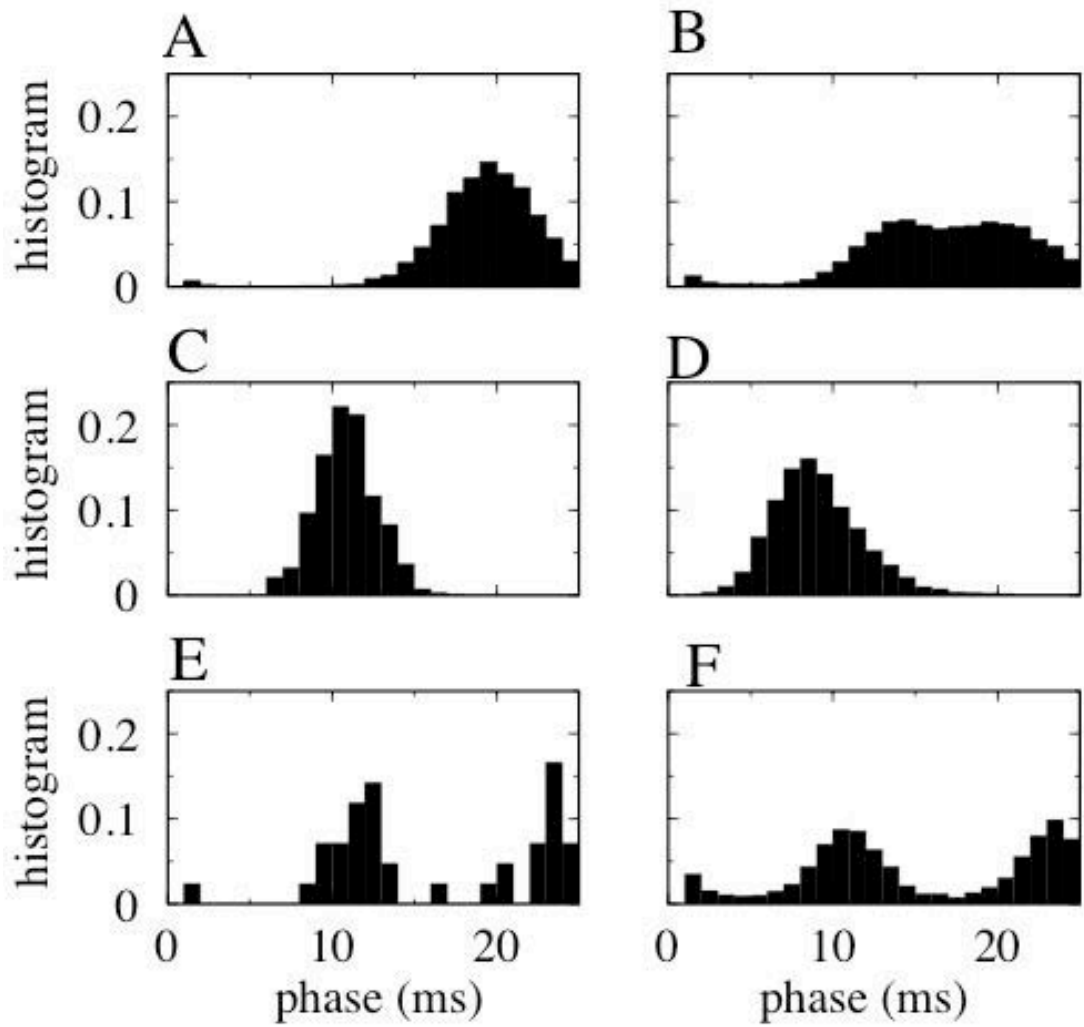

Figure 3

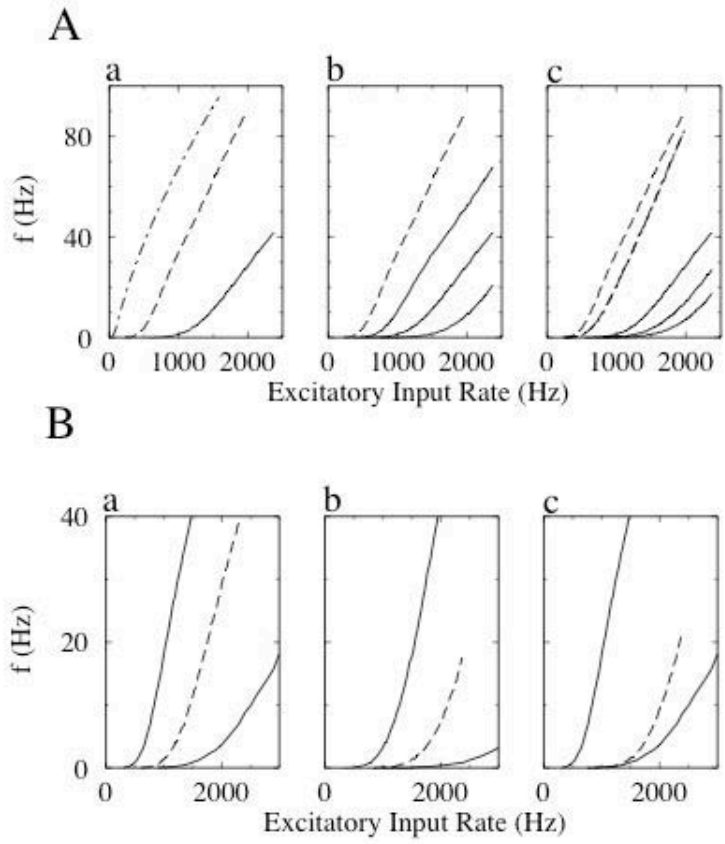

Figure 4

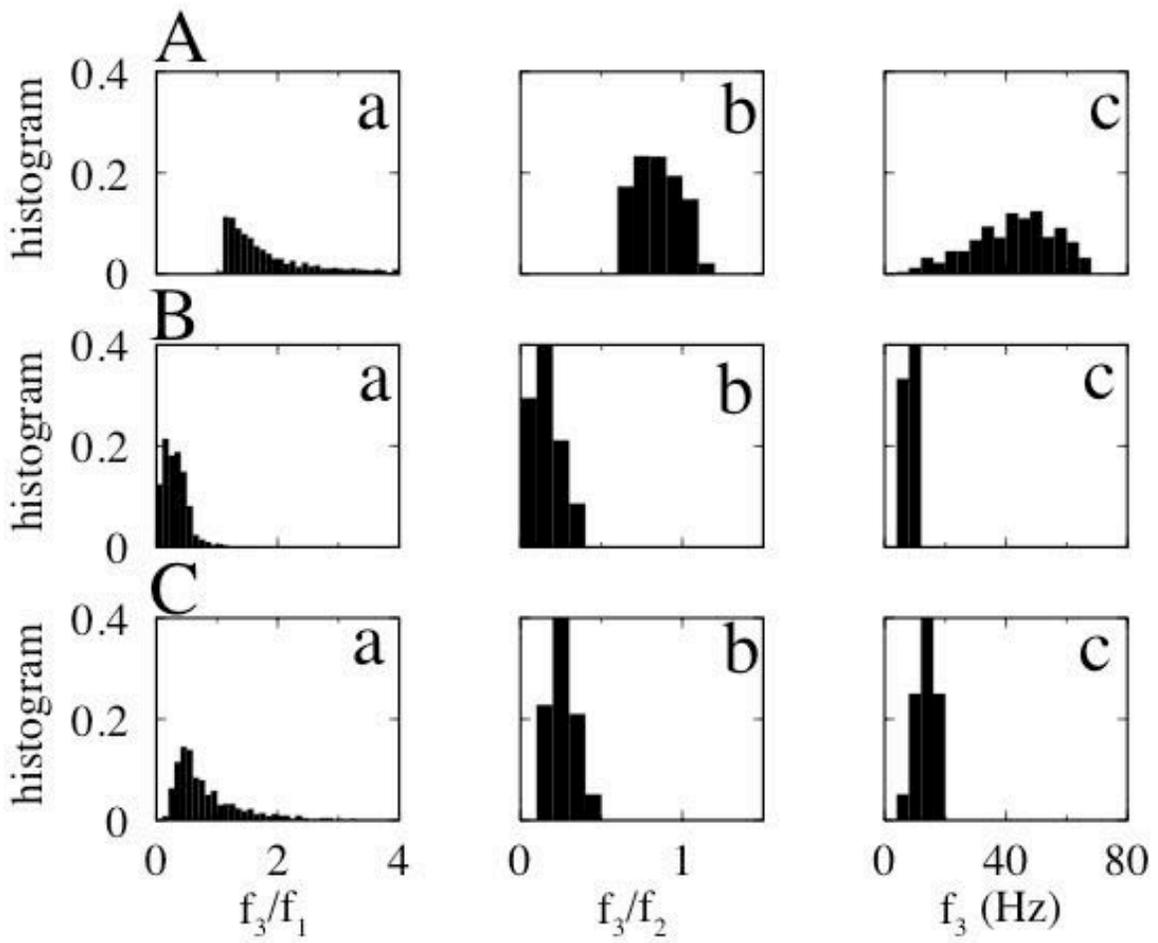

Figure 5

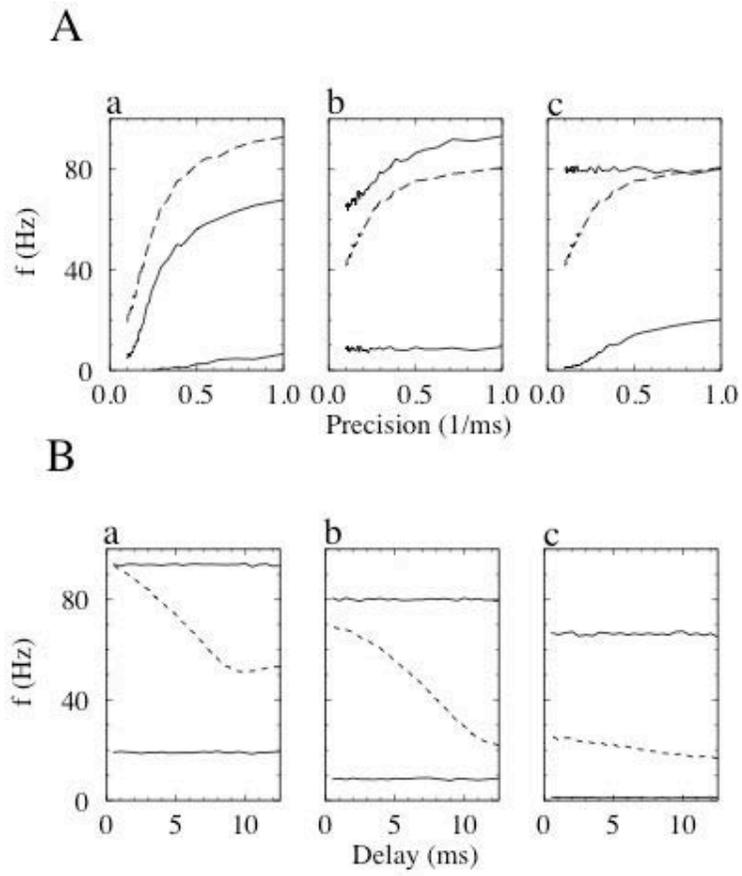

Figure 6

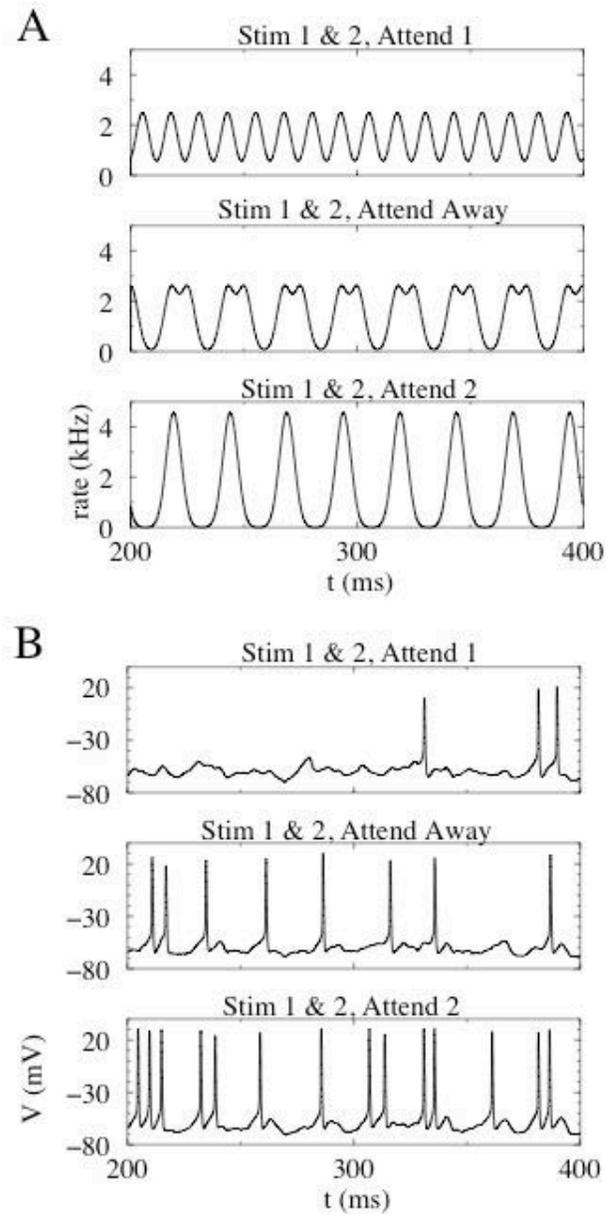

Figure 7